\documentstyle[multicol,epsfig,aps,prb]{revtex}
\begin{document}
\draft
\hyphenation{following}
\title{First principles calculation of structural and
 magnetic properties for Fe monolayers and bilayers on W(110)}
\author{X. Qian and W. H\"ubner}
\address{Max-Planck-Institut f\"ur Mikrostrukturphysik, Weinberg 2,
D-06120 Halle, Germany}
\date{\today}
\maketitle
\vskip 0.2cm
\begin{abstract}
Structure optimizations were performed for 1 and 2 monolayers (ML) of Fe on a 
5 ML W(110) substrate employing the all-electron full-potential linearized
augmented plane-wave (FP-LAPW) method. The magnetic moments were also obtained 
for the converged and optimized structures. We find significant contractions 
($\sim$ 10 $\%$) for both the Fe-W and the neighboring Fe-Fe interlayer spacings
compared to the corresponding bulk W-W and Fe-Fe interlayer spacings. Compared to the 
Fe bcc bulk moment of 2.2 $\mu_B$,  
the magnetic moment for the surface layer of Fe is enhanced (i) by 15$\%$ to 2.54
$\mu_B$ for 1 ML Fe/5 ML W(110), and (ii) by 29$\%$ to 2.84 $\mu_B$ for 2 ML Fe/5 ML
W(110).  The inner Fe layer for 2 ML Fe/5 ML W(110) has
a bulk-like moment of 2.3 $\mu_B$. These results agree well with previous 
experimental data. 
\end{abstract}

\begin{multicols}{2}
\section{Introduction}
\indent Magnetic thin films on metal substrates demonstrate fascinating phenomena
such as the preferential orientation of the magnetization normal to the film plane, 
enhanced low-temperature surface magnetization, and 
the pronounced effects of magnetism on the electrical conductivity. There has been
considerable
effort~\cite{Gradmann82,Przybylski87,Elmers88,Hong88,Przybylski89,Elmers89,Elmers90,Albrecht91,Albrecht92,Gradmann93,Elmers94,Elmers95,Bethge95,Sander96,Tober,Weber97,Fruchart,Sander98,Malzbender98,Hauschild98,Sander99}
in studying the atomic structures
and magnetic properties including magnetic moments and the orientation of the magnetic
easy axis for Fe thin films on W(110).  It is especially interesting to study these
properties for 1 and 2 monolayers (ML) of Fe on W(110) due to the pseudomorphic
layer-by-layer growth of the film when the Fe coverage $\theta$ is below 2 ML 
and to the possible magnetization reorientation
from in-plane to perpendicular for  1 $<\theta <$ 2. Both bulk Fe and W are bcc
structures with lattice constants of 2.86~\AA$\ $ and 3.165~\AA$\ $ respectively. 
It was found that Fe thin films grow pseudomorphically up to 1.2 ML~\cite{Sander96,Sander99} on the flat
W(110) surface and up to 1.8 ML~\cite{Gradmann82} on a vicinal surface.  
Significant structural relaxation in the vertical lattice spacings for the Fe
thin films is expected arising from the large lattice mismatch ($\sim$ 9 $\%$) between
the film and the W substrate. However, the exact amount of relaxation and the magnetic
moments for the Fe overlayers need to be clarified because of the conflicting results
between the experiments~\cite{Albrecht91,Albrecht92,Tober} and with previous theoretical
predictions~\cite{Hong88}. On account of
(i) the pseudomorphic growth of the Fe thin film, and (ii) the transitions of both 
the atomic structures and magnetic properties already in the ultrathin ($<$2 ML)
regime, it is feasible to employ the {\em ab initio} method to investigate these
properties.

\indent Earlier experimental work done by Albrecht {\em et al.}~\cite{Albrecht91,Albrecht92}
with low-energy electron 
diffraction (LEED) on 1 ML Fe on W(110) substrate showed that the Fe-W interlayer
spacing is contracted by 13$\%$ to 1.94~\AA$\ $ compared to the bulk W(110) interlayer  
spacing of 2.238~\AA. The magnetic moment for the top Fe layer is enhanced to 2.53 $\mu_B$
measured by Torsion Oscillation Magnetometry (TOM)~\cite{Gradmann93}.  However, recent work
done by Tober {\em et al}~\cite{Tober} using
Photo-Electron Diffraction (PED) for 1 ML Fe on W(110) yielded a Fe-W interlayer
spacing of 2.07~\AA, a relaxation of 7.2$\%$ only. Earlier {\em ab initio}
calculations by Hong {\em et al.}~\cite{Hong88} showed that the Fe-W interlayer distance
is dramatically
reduced by as much as 16$\%$ to 1.88~\AA. Because of this strong inward relaxation,
the magnetic moment of the overlayer Fe is only 2.18 $\mu_B$ which is very close to 
the bulk bcc Fe moment of 2.2 $\mu_B$. 
Recent calculations by Batirev {\em et al.}~\cite{Batirev98} showed that the Fe-W 
interlayer spacing is 
contracted by 3.1$\%$ with respect to the average theoretical bulk bcc(110) Fe and W
interlayer spacings. The magnetic moment for the Fe layer is 2.17 $\mu_B$ 
similar to the bulk value. Because of these differences between the experimental
results and also with the calculations, a detailed theoretical investigation is
presented to illuminate the incoherent data on structural and magnetic properties 
of Fe monolayers on W(110). 

\section{Method}
\indent It is well known that magnetic properties depend strongly upon the atomic
structures of the thin films.  Therefore it is necessary first to obtain an optimized
structure
for these systems. Three different slabs: (i) 5 ML W(110) clean substrate, (ii) 1 ML
pseudomorphic Fe overlayer on each side of 5 ML W(110) substrate, and (iii) 2 ML 
pseudomorphic Fe overlayers on each side of 5 ML W(110) substrate were studied. 
The schematic picture of 2 ML Fe on 5 ML W(110) is shown in Fig. 1. The bare
W(110) substrate was studied to test our theoretical accuracy since reliable
experimental results are available and theoretical calculations are abundant.

\indent These calculations were performed employing the WIEN97 code~\cite{Blaha90}.  
This program is based on the density-functional-theory (DFT)
and adopts the full-potential linearized augmented plane-wave (FP-LAPW) method.
It has the additional capability of computing atomic forces~\cite{Kohler95,Kohler96,Kohler97}
which makes the
structure optimization much more efficient compared to the total energy only
calculations.  The FP-LAPW method adopts different representations for wave 
functions, charge density and potential inside the muffin-tin sphere 
and in the interstitial region. The spherical harmonics were expanded
up to  l=10 inside the muffin-tin and to l=4 for the interstitial in the present
calculations. Spin-polarized calculations were carried out in order to determine the
magnetic properties. Spin-polarization was implemented in the WIEN97 code adopting
Local Spin-Density Approximation (LSDA) with two separate spin densities. Two
sets of Kohn-Sham (KS) orbitals for the two spin components were obtained, and two sets
of KS single particle equations were solved.  The scalar relativistic 
calculations including the \emph{velocity} and the \emph{Darwin} terms were adopted
for valence electrons.  Spin-orbit coupling for the valence electrons was not included
for the present calculations. The shallow 5p states were treated as semi-core, i.e. as
local orbitals, thereby ensuring the flexibility of the basis functions to closely
represent these low-lying p orbitals.  Pulay corrections~\cite{Kohler96,Pulay} to the
Hellmann-Feynman 
forces were calculated which makes the structure optimization highly accurate. 
The improved tetrahedron method~\cite{Blochl} was used for the integrations.

\indent Fig. 2 shows the unit cells for the calculations. The structure optimization
for the slabs was done by giving an initial guess
of the interlayer spacings based on the optimized structure of Fe/Mo(110).  The
direction and degree of relaxation for the
vertical interlayer spacings depend on the magnitude and sign of the forces
present.  The in-plane lattice constant for the slab was fixed and
taken from the bulk calculations and will be described later. This is due to 
the in-plane two-dimensional translational invariance and the fact that there
is only one atom on each layer in the unit cell.  As described in our previous 
paper~\cite{qian}, eight vacuum layers were incorporated in the supercell to separate
the slabs in order to minimize any Coulomb and exchange interactions.
Furthermore, slabs are symmetric with respect to the central substrate layer
to avoid any charge accumulation on the surfaces. 
Thus the contribution to the total energy from the electric-dipole interaction
between the supercells is negligible compared to the contributions from 
within the supercell.  In addition, only real wave functions 
are needed for the calculations because of the presence of inversion symmetry.  
The Fe layers on each surface are ferromagnetically coupled.  The spin-polarized
calculations were applied.

\indent In these calculations, Generalized Gradient Approximation 
(GGA)~\cite{Perdew92G} exchange
potential and scalar-relativistic treatment were used in agreement with our earlier
calculations on Mo substrate. Generally speaking, we did not find any significant 
improvement of GGA exchange potential over LSDA potential.  Following the
procedure described previously~\cite{qian}, 
the theoretical bulk W lattice constant was determined to be 3.205~\AA , 1.3$\%$
larger than the experimental value of 3.165~\AA. It is known that GGA 
corrects overbinding, but sometimes leads to an excessive increase in the lattice
parameter for heavy atoms such as W. Nevertheless this theoretical value was
used as the in-plane lattice spacing in our subsequent slab calculations. 
The theoretical bulk Fe lattice constant was found to be 2.834~\AA, 0.9$\%$ smaller
than the experimental result of 2.86~\AA.
The muffin-tin radii were chosen to be 1.27~\AA$\ $and 1.164~\AA$\ $for W and Fe atoms
respectively in the slab unit cells. Convergence was achieved when the 
total energy and charge differences between two consecutive 
iterations are less than 5x$10^{-5}$ Ry and 1x$10^{-4}$ e$/$(a.u.)$^3$
respectively.  The structure optimizations were done when the force on each atom is
less than 1 mRy/a.u.. The magnetic moments were calculated as the differences between the
spin-up charge and spin-down charge for these converged results. Orbital magnetic moment is
not included in our calculations due to the absence of spin-orbit coupling for the 
valence electrons.  Moreover, it was previously estimated to be around 0.1 $\mu_B$~\cite{Bruno89} only.
The numbers of {\bf k} points in the two-dimensional meshes are
20 x 20 for 5 ML W(110), 21 x 21  for 1 ML Fe on 5 ML W(110), and 
22 x 22 for 2 ML Fe on 5 ML W(110).  The numbers of {\bf k} points
in the irreducible part of the Brillouin zone (IBZ) (1/4 of BZ) are 110, 121, 
and 132 respectively. The plane-wave cut-offs (corresponding to the largest k-vector in
the plane-wave basis expansion) are 16.7, 15.3, and 13.2 Rys for the three slabs 
respectively with 0, 1 and 2 ML Fe coverage. The kinetic
energy cut-offs (corresponding to the largest reciprocal-space vector for the potential
expansion) are 196 Ry for all three slabs.

\section{Results and Discussion}

\indent The structural and magnetic results are exhibited in Tables I and II respectively.
For the 5 ML W(110) clean substrate, we find that the top W-W interlayer spacing is
contracted by 4.1$\%$ to 2.173~\AA$\ $ from the theoretical bulk W-W interlayer spacing
of 2.266~\AA$\ $ in the (110) plane.  This result is in good agreement with the previous
FP-LAPW calculations~\cite{Arnold} in which the same amount of contraction (4.1$\%$) was found for the
top W-W interlayer spacing with a 5 ML W(110) slab. A 3.6$\%$ downward relaxation was found
for the top layer with a 9 ML W(110) slab. Our result is in disagreement with a recent
calculation~\cite{Batirev98}.  However, in that study, only three substrate W(110) layers were employed.
The recent LEED experiment~\cite{Arnold} yielded a 
contraction of 3.1$\%$ with an error bar of 0.6$\%$.  In addition to our agreement with 
previous theoretical and experimental data, our present result is also quite similar to
the relaxation found for a 5 ML Mo(110) slab published earlier~\cite{qian}. 
Further, we find that the second W-W interlayer distance is also slightly contracted 
by 0.4$\%$ to 2.258~\AA.  This again agrees well with earlier calculations~\cite{Arnold}
in which a 0.2$\%$ contraction was found for the 5 ML W(110) slab. 

\indent The clean W(110) substrate is non-magnetic. The density-of-state
(DOS) plot is shown in Fig. 3. Only {\em d}-partial DOS (PDOS) of spin-down are shown since 
they are identical to the spin-up DOS. The inner
W layer (W(S-2)) {\em d}-PDOS closely resembles the bulk bcc W {\em d}-PDOS. The
surface layer (W(S)) {\em d}-PDOS has a higher number of states at the Fermi-level, almost
double that of the W(S-2), i.e. a less pronounced gap between the two sub-bands.
 
\indent For the slab of 1 ML Fe on each side of 5 ML W(110) substrate, we find a significant
relaxation for the Fe-W interlayer spacing (Table I) very similar to the case of 1 ML
Fe/5 ML Mo(110) as shown in our previous work~\cite{qian}. The Fe-W interlayer has a downward
relaxation 
of 12.9$\%$ compared to the bulk W-W interlayer distance.  It is in excellent agreement with
the LEED experiment by Albrecht {\em et al.}~\cite{Albrecht91,Albrecht92} in which a 13$\%$
contraction was found compared
to the bulk experimental W-W interlayer distance. The recent PED experiment, however, yielded
a Fe-W distance of 2.07~\AA (7.2$\%$ contraction only) which corresponds to the bond length from the hard sphere's model.
The earlier calculations done by Hong {\em et al.}~\cite{Hong88}
showed a much larger downward relaxation of 16$\%$ employing the FP-LAPW method. However, in 
their earlier calculations, not all the atoms were allowed to relax at the same time 
since it was not possible to compute the force on each atom. In addition to the Fe-W distance,
our present calculations show that the neighboring W-W interlayer spacing is reduced slightly
by 0.1$\%$. However our earlier results on Mo(110) show a small expansion for the neighboring
Mo-Mo interlayer spacing contrary to the W case here. 

\indent  The magnetic moment for the surface layer of Fe is found to be 2.54 $\mu_B$ without
orbital moment contribution, an enhancement of 15$\%$ over the bulk magnetic moment of
2.2 $\mu_B$ for bcc Fe. However, it is reduced by 29$\%$ compared to the moment of 3.3 $\mu_B$ for 
the Fe(110) free-standing monolayer with the same in-plane lattice parameter. In addition,
our results show that the neighboring W layer acquires a small moment of 0.1 $\mu_B$.
It is antiferromagnetically coupled to the Fe overlayer.  The 
Torsion Oscillation Magnetometry(TOM) experiment done by Gradmann and
coworkers~\cite{Gradmann93} yielded a moment of 2.53 $\mu_B$ for the overlayer Fe.  
Since the orbital moment and the induced substrate moment are both around 0.1 $\mu_B$ and
opposite in sign, the theoretical spin moment we obtained for Fe overlayer agrees very well
with the TOM experiment since TOM measures the total moment.
 The earlier calculations done by Hong
and coworkers~\cite{Hong88} showed no enhancement of the moment over the bulk value. It is
probably due to the fact that their calculations yield a significant reduction of the
Fe-W interlayer distance.

\indent Spin-down and spin-up {\em d}-PDOS for both the surface Fe layer and the neighboring
W layers are plotted in Figs. 4 and 5 respectively.  The {\em d}-PDOS of both spins
for the inner W(S-2) layer are very similar to the central W layer of the previous case where there is no Fe 
overlayer. The spin-down {\em d}-PDOS of the interfacial W(S-1) layer resembles
the one of the inner W(S-2) layer. However, there are noticeable changes for the
spin-up component especially when close to the Fermi surface.  The small moment of the W(S-1) layer
is due to this change of {\em d}-PDOS. The overlayer Fe {\em d}-PDOS are very different to 
their corresponding bulk ones as shown in Figs. 6 and 7, especially for the spin-up component.
Figs. 6 and 7 will be discussed a little later.

\indent For the slab of 2 ML Fe on each side of 5 ML W(110) substrate, we find both the
Fe-Fe and Fe-W interlayer spacings are contracted dramatically (see Table I). The Fe-Fe
interlayer distance
is reduced by 11.9 $\%$~\cite{standard} from the theoretical bulk Fe value of 2.004~\AA$\ $ to 1.766~\AA. The
Fe-W interlayer spacing is contracted by 10.6$\%$ compared to the bulk W-W interlayer distance.
The percentage of the contractions are also very close to the Fe(110)/Mo(110) case~\cite{qian}.
Albrecht {\em et al.}~\cite{Albrecht91,Albrecht92}
found a 10$\%$ downward relaxation for the Fe-Fe interlayer
spacing compared to the bulk Fe value. Again it is in excellent agreement with our findings.
Our calculations show a slight expansion $<$ 0.2$\%$ for the inner W-W interlayer spacings.  

\indent The magnetic moment for the surface layer of Fe is found to be 2.84 $\mu_B$, an 
enhancement of 29$\%$ over the bulk value of 2.2 $\mu_B$.  It is still smaller than the moment 
of the Fe(110) free-standing monolayer. However, compared to the 1 ML Fe/5 ML W(110) case, the moment
for the top Fe layer is increased from 2.54 $\mu_B$ to 2.84 $\mu_B$. This is probably due to
the strong hybridization of Fe {\em d}-orbitals with the ones of interfacial W layer thereby
reducing the moment of the Fe layer. The second Fe layer, i.e. the interfacial Fe layer 
has a moment of 2.3 $\mu_B$ already very close to the bulk value. Like the previous case, the
neighboring substrate layer also acquires a small moment of 0.1 $\mu_B$ and is 
antiferromagnetically coupled to the Fe overlayers. 

\indent The {\em d}-PDOS for the surface and interfacial Fe layers together with the ones of
 Fe bcc bulk are plotted in Fig. 6 (spin-down) and Fig. 7 (spin-up) for comparison. Basically
 the {\em d}-PDOS of the second layer of Fe are already close to the bulk ones. Consequently
 its magnetic moment is also approaching the bulk value.  The {\em d}-PDOS of the surface Fe
 layer are different from the bulk ones particularly for the spin-up component and when 
 close to the Fermi-surface for the spin-down component.
 
\section{Summary}
\indent The present FP-LAPW calculations resolves the discrepancies between previous
experimental data and with earlier theoretical results on the atomic structure and 
magnetic moment of 1 ML Fe/W(110).  The Fe-W interlayer spacing is significantly 
contracted by  as much as $\sim$13$\%$ compared to the bulk W-W interlayer spacing. 
The magnetic moment
of the overlayer Fe is greatly enhanced compared to the bulk moment of bcc Fe due to
the lower coordination number, but it is reduced compared to the Fe(110) free-standing
monolayer because of the presence of the substrate.

\section{Acknowledgement}
\indent The authors would like to thank A.J. Freeman, U. Gradmann, J. Kirschner,
and D. Sander for stimulating discussions.

\end{multicols}
\newpage
\begin{table}
\begin{center}
\begin {tabular}{|p{3.8cm}|p{3.2cm}|p{2.5cm}|p{2.5cm}|p{2.5cm}|}
   & d($Fe_2$-$Fe_1)$ & d($W_2$-$Fe_1$) & d($W_1$-$W_2$) & d($W_2$-$W_3$)\\\hline
5 ML W(110) & & & 2.173(-4.1\%)& 2.258(-0.4\%)\\
1 ML Fe/W(110)  & & 1.974(-12.9\%) & 2.263(-0.1\%) & 2.251(-0.7\%)\\
2 ML Fe/W(110) & 1.766(-11.9\%)~\cite{standard} & 2.026(-10.6\%) & 2.267 (0.03\%) & 2.272(0.2\%)\\
\hline
W(110) (Exp.)~\cite{Arnold} & & & 2.169(-3.1\%) & \\
Fe/W(110)(Exp.)~\cite{Albrecht92} & 1.82(-10\%) &&&\\
Fe/W(110)(Exp.)~\cite{Albrecht91} && 1.94(-13\%) & & \\
Fe/W(110)(Exp.)~\cite{Tober} & & 2.07(-7.2\%) & 2.28(2.2\%) &\\
1 ML Fe/W(110)~\cite{Hong88} & &1.88(-16\%)& & \\
\end{tabular}
\vskip 1cm
\caption{Structural results (The layer spacings are given in~\AA. The relative changes as compared to the bulk W layer 
spacing are given in parentheses. The percentage of Fe-Fe contraction is relative to the bulk Fe-Fe
interlayer spacing.)}
\end{center}
\end{table}

\newpage
\normalfont
\begin{table}
\begin{center}
\begin {tabular}{|p{4.8cm}|p{3.8cm}|p{3.8cm}|}
& 1 ML Fe/W(110) ($\mu_B$) & 2 ML Fe/W(110) ($\mu_B$)\\\hline
Fe(2) & & 2.850(2.844)\\
Fe(1) & 2.561(2.536) & 2.315(2.308)\\
W(1) & -0.085 & -0.104\\
W(2) & -0.000 & -0.004 \\
W(3) & -0.000 & -0.006 \\
Interstitial & -0.055 & -0.04 \\
\hline
Fe(2)(Exp.)Fe/W(110))~\cite{Gradmann93} & & 2.77 \\
Fe(1)(Exp.)Fe/W(110))~\cite{Gradmann93} & 2.53 & \\
\end{tabular}
\vskip 1cm
\caption{Magnetic moments (values in the parentheses are obtained by
 adding interstitial contributions).}
\end{center}
\end{table}

\newpage
\begin{figure}
\begin{center}
\epsfig{file=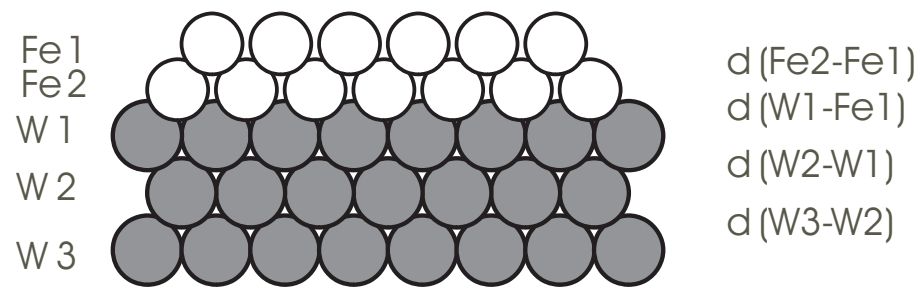, width=15cm, angle=0}
\vskip 1cm
\caption{Schematic picture of 2 ML Fe on 5 ML W(110)(upper half of the slab only).}
\end{center}
\end{figure}

\newpage
\vskip -0.2in
\begin{figure}
\begin{center}
\epsfig{file=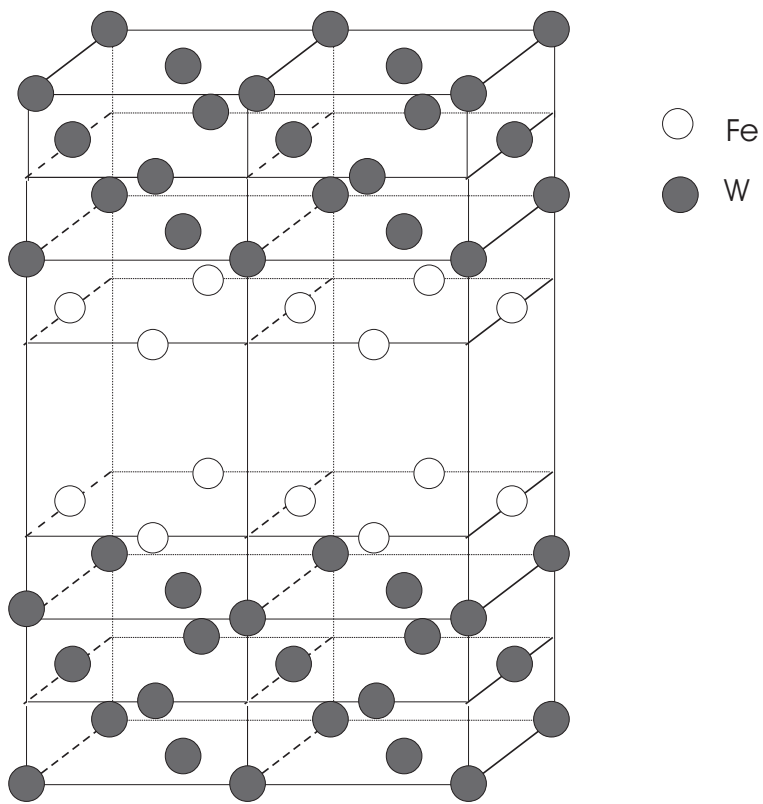, width=10cm,angle=0}
\vskip 1cm
\caption{Unit cells of the 1 ML Fe on each side of 5 ML W(110).}
\end{center}
\end{figure}

\newpage
\vskip -0.2in
\begin{figure}
\begin{center}
\psfig{file=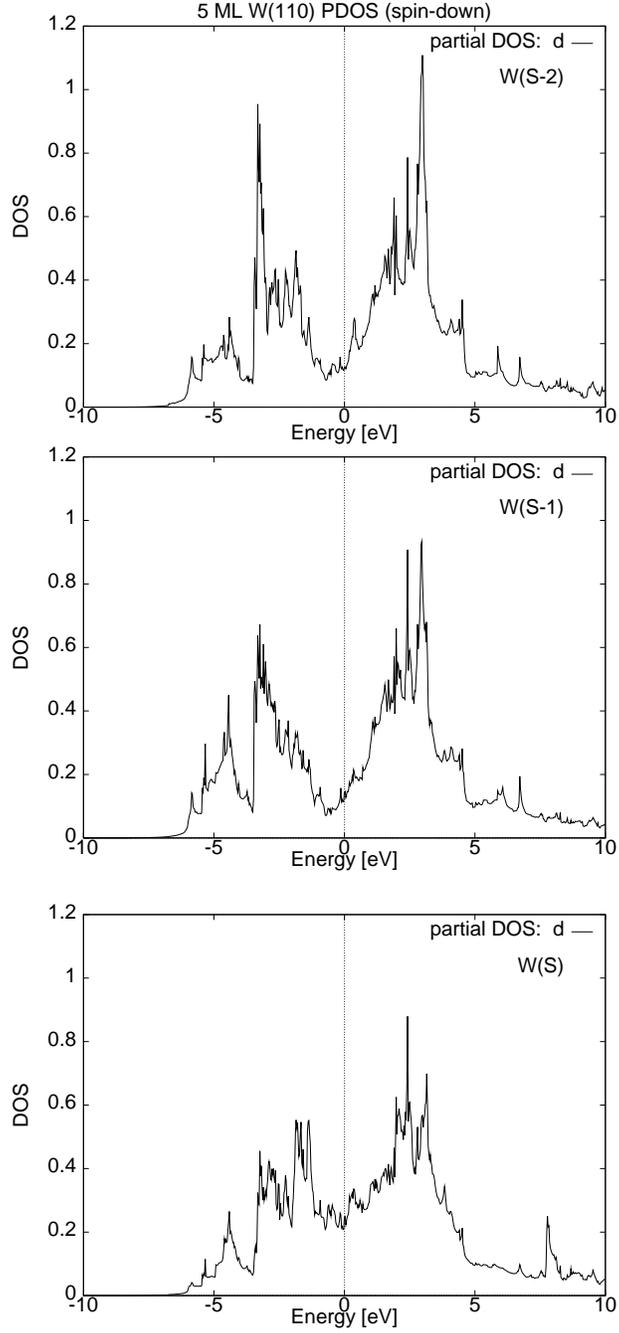, width=15cm,angle=0}
\caption{Spin-down partial-{\em d} density-of-states({\em d}-PDOS) for 5 ML W(110) clean substrate. S represents the 
surface layer, S-1 the layer next to the surface layer, and S-2 the central layer.}
\end{center}
\end{figure}

\newpage
\vskip -0.2in
\begin{figure}
\begin{center}
\psfig{file=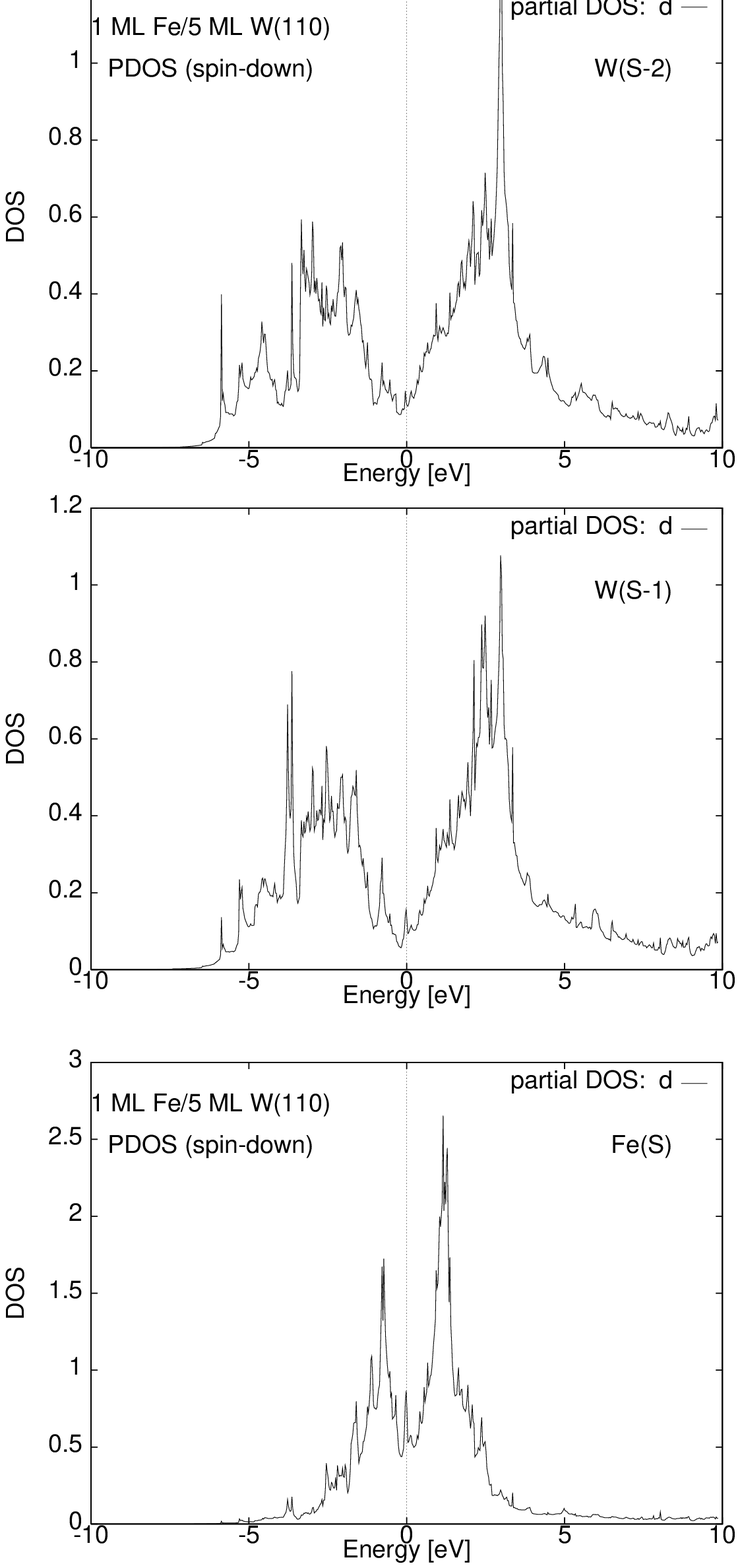, width=15cm,angle=0}
\caption{Spin-down partial-{\em d} density-of-states({\em d}-PDOS) for 1 ML Fe/5 ML W(110).}
\end{center}
\end{figure}

\newpage
\vskip -0.2in
\begin{figure}
\begin{center}
\psfig{file=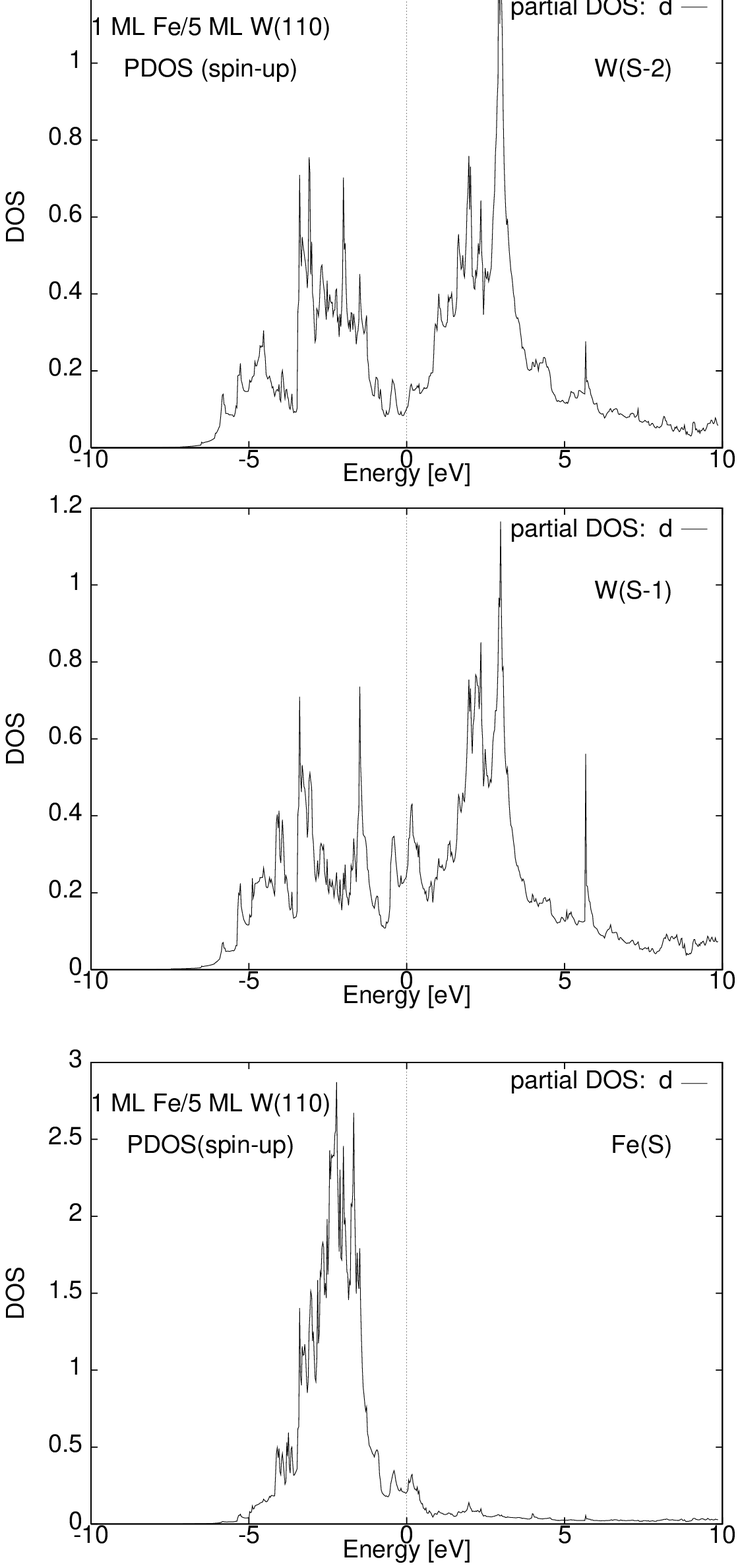, width=15cm,angle=0}
\caption{Spin-up partial-{\em d} density-of-states({\em d}-PDOS) for 1 ML Fe/5 ML W(110).}
\end{center}
\end{figure}

\newpage
\vskip -0.2in
\begin{figure}
\begin{center}
\epsfig{file=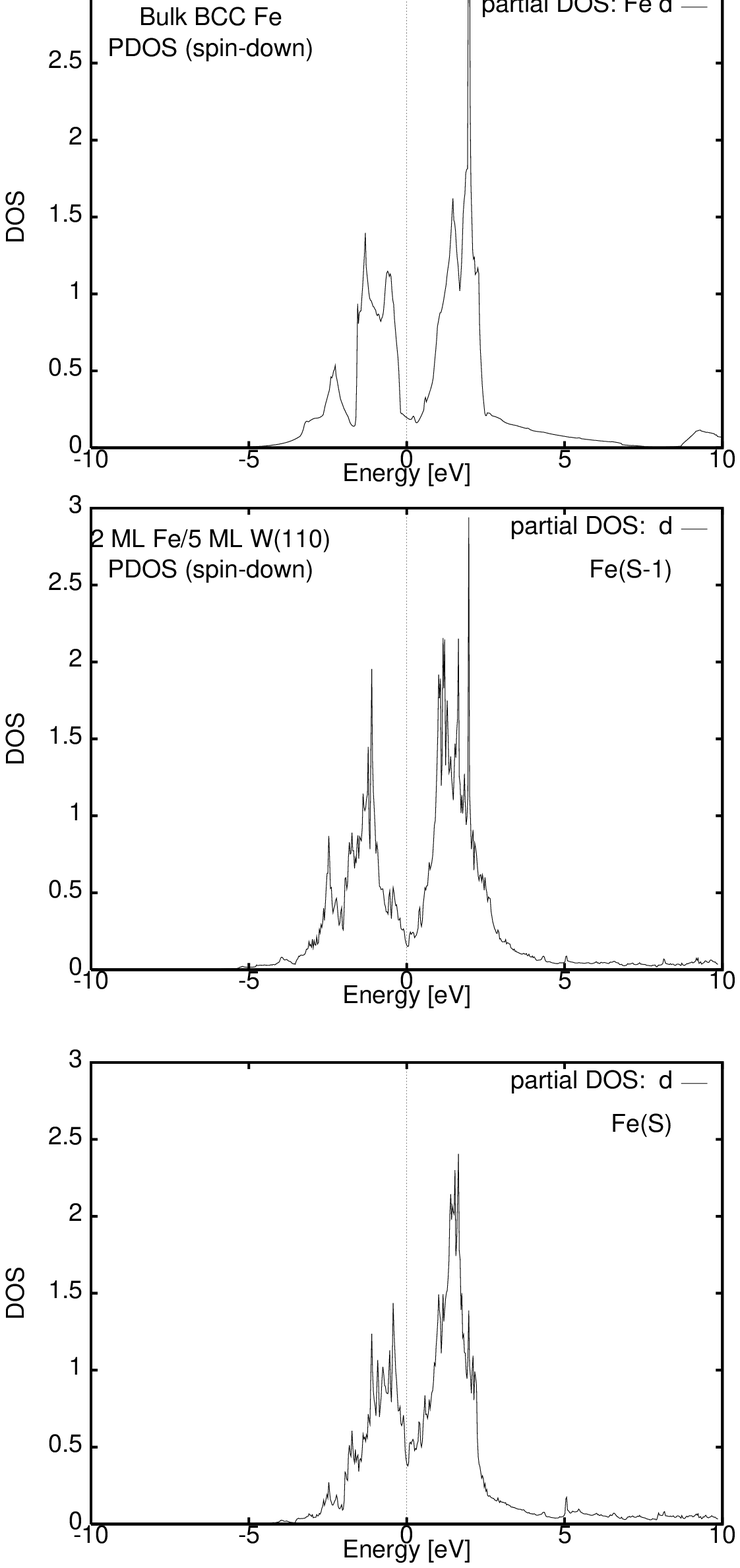, width=15cm,angle=0}
\caption{Comparison of the Fe spin-down partial-{\em d} density-of-states({\em d}-PDOS) for
 2 ML Fe/5 ML W(110) with the corresponding bulk bcc Fe one.}
\end{center}
\end{figure}

\newpage
\vskip -0.2in
\begin{figure}
\begin{center}
\psfig{file=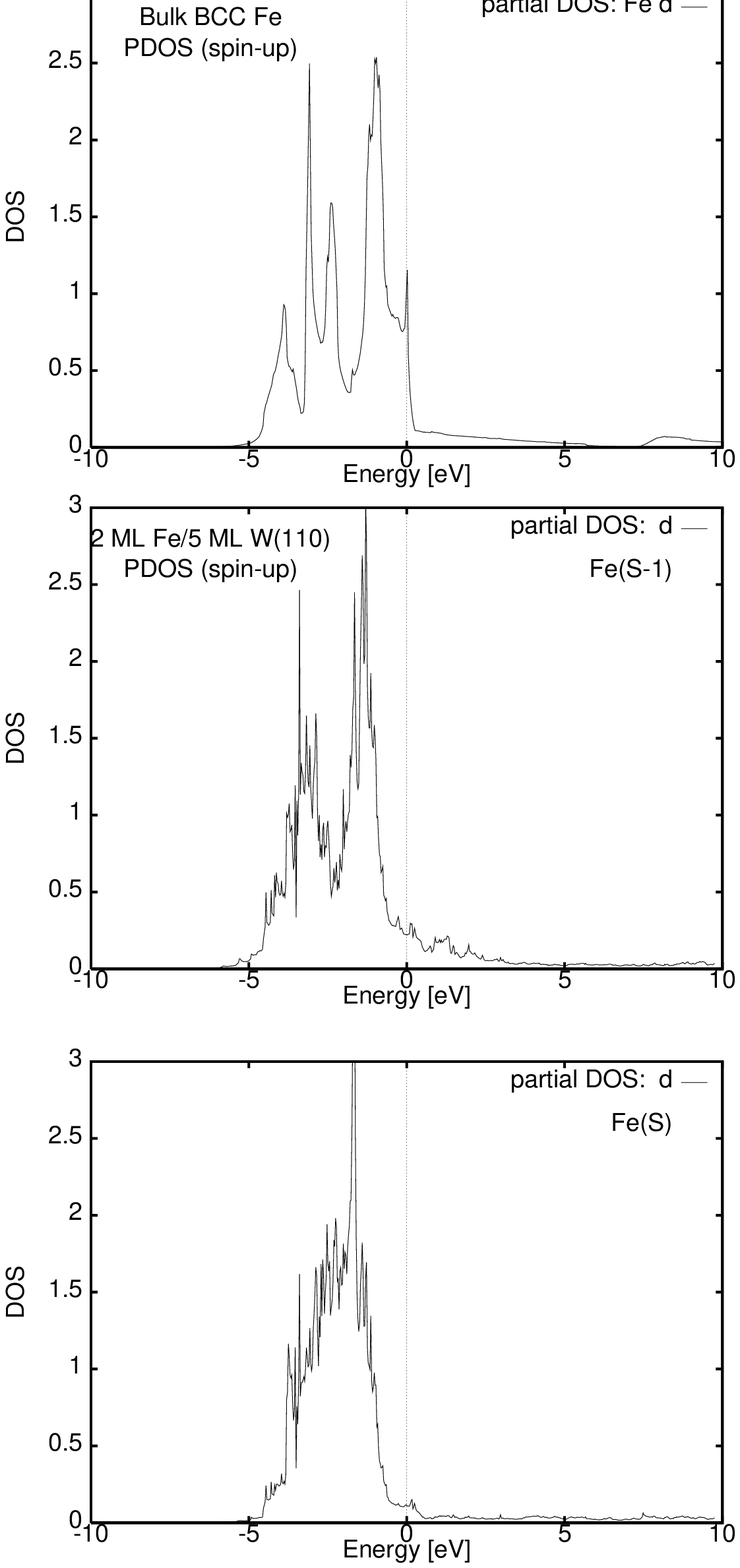, width=15cm,angle=0}
\caption{Comparison of the Fe spin-up partial-{\em d} density-of-states({\em d}-PDOS) for
 2 ML Fe/5 ML W(110) with the corresponding bulk bcc Fe one.}
\end{center}
\end{figure}

\begin{references}
\bibitem{standard} The Fe-Fe interlayer distance is contracted by 22.1 $\%$ with respect
to the theoretical bulk W-W interlayer spacing.
\end{references}
\end{document}